\documentclass[aip,jcp,reprint,twocolumn,showpacs]{revtex4}

\pdfoutput=1

\usepackage{amsmath,amssymb}
\usepackage{bm}
\usepackage{graphicx}

\usepackage{color}

\usepackage{ulem}
\definecolor{blau}{rgb}{0.2,0.2,1.}

\begin{document}

{\it\small
\noindent Journal Reference: J.~Chem.~Phys.~\textbf{141}, 194907 (2014)\\
URL: http://scitation.aip.org/content/aip/journal/jcp/141/19/10.1063/1.4901275  \\
DOI: 10.1063/1.4901275   \\[1.cm]
}

$\mbox{\hspace{1.cm}}$
\vspace{.9cm}

\title{
Bridging from particle to macroscopic scales in uniaxial magnetic gels
}

\author{Andreas M.\ Menzel}
\email{menzel@thphy.uni-duesseldorf.de}
\affiliation{Institut f\"ur Theoretische Physik II: Weiche Materie, Heinrich-Heine-Universit\"at D\"usseldorf, D-40225 D\"usseldorf, Germany
}

\date{\today}

\begin{abstract}
Connecting the different length scales of characterization is an important, but often very tedious task for soft matter systems. Here we carry out such a procedure for the theoretical description of anisotropic uniaxial magnetic gels. The so-far undetermined material parameters in a symmetry-based macroscopic hydrodynamic-like description are determined starting from a simplified mesoscopic particle-resolved model. This mesoscopic approach considers chain-like aggregates of magnetic particles embedded in an elastic matrix. Our procedure provides an illustrative background to the formal symmetry-based macroscopic description. 
There are presently other activities to connect such mesoscopic models as ours with more microscopic polymer-resolved approaches; together with these activities, our study complements a first attempt of scale-bridging from the microscopic to the macroscopic level in the characterization of magnetic gels. 
\end{abstract}



\maketitle


\section{Introduction} \label{sec:Introduction}

Ferrogels and magnetic elastomers, both of which we simply refer to as magnetic gels in the following, consist of colloidal magnetic particles embedded in a crosslinked polymer matrix \cite{filipcsei2007magnetic}. They are therefore composite materials that combine the features of two different prominent components: on the one hand, the properties of magnetic colloidal fluids \cite{rosensweig1985ferrohydrodynamics,odenbach2002ferrofluids,odenbach2003ferrofluids,odenbach2003magnetoviscous, huke2004magnetic,odenbach2004recent,fischer2005brownian,ilg2005structure, holm2005structure,klapp2005dipolar,gollwitzer2007surface, jordanovic2008field,vicente2011magnetorheological}, which for instance allow a reversible adjustment of their viscosity by external magnetic fields \cite{rosensweig1969viscosity,mctague1969magnetoviscosity, thurm2002magnetoviscous,thurm2003particle, odenbach2003ferrofluids,odenbach2004recent,ilg2005anisotropy,pop2006investigation}; and, on the other hand, the elastic behavior of conventional rubbers \cite{treloar1975physics,strobl1997physics}. 

The combination of these different material properties and their coupling in one substance opens the way to new applications. For example, by an external magnetic field the elastic moduli of magnetic gels can be reversibly adjusted from outside \cite{deng2006development,filipcsei2007magnetic,stepanov2007effect, bose2009magnetorheological,evans2012highly, borin2013tuning}. 
Consequently, the possibility of constructing externally tunable damping devices \cite{sun2008study} or vibration absorbers \cite{deng2006development} has been outlined. Also, the deformations of the materials in external magnetic fields have been investigated \cite{zrinyi1996deformation,guan2008magnetostrictive, stolbov2011modelling,gong2012full, zubarev2013magnetodeformation,allahyarov2014magnetomechanical} and their use as soft actuators \cite{zimmermann2006modelling,filipcsei2007magnetic} and as magnetic sensors \cite{szabo1998shape,ramanujan2006mechanical, liu2006magnetic} has been pointed out. 

There have been recent attempts to increase the magneto-mechanical coupling in these materials \cite{frickel2009functional,frickel2011magneto,messing2011cobalt}. For this purpose, the polymer chains of the embedding elastic matrix were directly chemically anchored on the surfaces of the magnetic particles. In this way, rotational torques on the magnetic particles are directly transmitted to the surrounding polymer network \cite{weeber2012deformation}. Thus the particles cannot be reoriented independently of their environment, which can be interpreted in terms of an orientational memory \cite{annunziata2013hardening}. 

A complete theoretical characterization of magnetic gels is difficult to achieve, even for simulation approaches. The reason are the different length scales that are simultaneously addressed. An example is given by considering the elastic deformation of a block of material. The overall deformation occurs on a macroscopic length scale and is characterized, for instance, by the magnitude of the elastic moduli. However, the nature of the mechanical response and other characteristic material properties can strongly depend on the spatial distribution of the magnetic particles within the sample 
\cite{filipcsei2007magnetic,ivaneyko2011magneto,wood2011modeling,camp2011effects, stolbov2011modelling,ivaneyko2012effects,han2013field, ivaneyko2014mechanical,tarama2014tunable,pessot2014structural}. Typical particle diameters range from nano- to micrometers and thus involve an intermediate, i.e. mesoscopic length scale. Finally, for rubbery substances, the fact that the materials are elastic mainly results from the entropic properties of the single crosslinked polymer chains and is therefore of molecular microscopic origin \cite{strobl1997physics}. 

Here, we confine ourselves to the theoretical characterization of anisotropic uniaxial magnetic gels. Such materials can be prepared by applying a strong external magnetic field during synthesis. Under such a situation, the magnetic particles tend to form chain-like aggregates \cite{collin2003frozen,varga2003smart,filipcsei2007magnetic,gunther2012xray, borbath2012xmuct,gundermann2013comparison} that can span the whole sample \cite{gunther2012xray}. After the polymer matrix has been crosslinked, the positions of the particles remain relatively fixed within the embedding polymer network. Thus the structural {arrangement into} chain-like aggregates persists even when the external magnetic field applied during synthesis is switched off. We can say that a positional memory of the particle locations is imprinted into the materials. 

Based on symmetry considerations, a macroscopic hydrodynamic-like theoretical characterization of these materials was derived already a decade ago \cite{bohlius2004macroscopic}. Per construction, this theory contains several material parameters, often referred to as phenomenological coefficients, that remain undetermined in the symmetry-based approach. These parameters must either be measured in an experiment, or they must be derived from a more microscopic approach. So far, to the best of our knowledge, neither of these tasks has yet been completed up to the present for uniaxial magnetic gels. 

In this paper, we address the second of these two tasks. We start from a simple mesoscopic model. Our goal is to determine expressions for the material parameters of the static part of the macroscopic theory in terms of the mesoscopic magnetic particle parameters. The underlying model is mesoscopic in the sense that it resolves the colloidal magnetic particle level. However, it does not resolve the individual polymer chains, which here we refer to as the microscopic level. Instead, the embedding polymer matrix is treated as an elastic continuum. In this way, we achieve a scale bridging from the mesoscopic to the macroscopic level in a (quasi-)static theoretical approach. Apart from the explicit expressions for the material parameters, this procedure provides us with an illustrative background to the formal macroscopic description. 

The structure of the paper is as follows. In the next section, we repeat the material-specific static part of the macroscopic theory, i.e.\ of the generalized energy density. The background of the different terms and variables is explained. After that, in section \ref{mesomodel}, we introduce our simple mesoscopic model of chain-like magnetic aggregates embedded in an elastic matrix. 
In the scale-bridging part, namely in section \ref{scalebridging}, we calculate the macroscopic material parameters in terms of the parameters of the mesoscopic model. 
{Magnetostrictive contributions in the context of nonlinear strain deformations are addressed in section \ref{magnetostriction}.} 
We briefly discuss our results in section \ref{discussion}. 
Finally, in the last section, we provide a short summary and include several remarks about future tasks. 

\section{Macroscopic theory}\label{macrotheory}

Here we briefly review the part of the macroscopic theory on uniaxial magnetic gels \cite{bohlius2004macroscopic} that we will concentrate on in the following. We restrict our approach to the static part of the theory, which is controlled by the generalized energy density $F$. Minimizing the corresponding generalized energy $\mathcal{F}=\int_VF\,d^3r$ of the system, with $V$ the volume of the material, leads to the actual static state of the system. For slowly varying external fields this procedure can also reproduce quasistatic material behavior. ``Slow'' in this case is defined relatively to all relevant time scales of the material. 

{
For simplicity, we consider spatially homogeneous situations. Moreover, only the effect of macroscopic variables that are not already present in the hydrodynamic description of normal fluids will be investigated. 
Our first step is to identify the symmetry properties. 
As we mentioned above, we will concentrate on uniaxial materials that feature aligned chain-like aggregates of magnetic particles embedded in the polymer matrix \cite{collin2003frozen,varga2003smart,filipcsei2007magnetic,gunther2012xray, borbath2012xmuct,gundermann2013comparison}. Naturally, the orientation of the macroscopic anisotropy axis coincides with the orientation of the chain axes. In the absence of an oblique or perpendicular external magnetic field and not considering interactions between different chains, the magnetization direction of each magnetic chain will be along its axis. Under these conditions, the ground-state orientation of the anisotropy axis is thus also parallel to the directions of the chain magnetizations. We refer to this average orientation of the magnetic moments as $\mathbf{\tilde{m}}$, where the tilde will distinguish it from the particle magnetic moment.} {Two situations are possible and will be specified further in the next section: either $\mathbf{\tilde{m}}$ is a genuinely polar vector, which identifies and distinguishes between a ``head'' and a ``tail'' of the system; or $\mathbf{\tilde{m}}$ only identifies the orientation of the anisotropy axis, while the system is symmetric under the transformation $\mathbf{\tilde{m}}\leftrightarrow -\mathbf{\tilde{m}}$.} In both cases, $\mathbf{\tilde{m}}$ serves to describe the orientational state of the magnetic component. 

In addition to that, the elastic component is characterized by its elastic deformations. We denote the positions of the material elements of the sample in the undistorted state by $\mathbf{r}$ and in the distorted state by $\mathbf{r}'$. Following linear elasticity theory \cite{landau1986elasticity}, we define the displacement field $\mathbf{u}=\mathbf{r'}-\mathbf{r}$. The transformation of infinitesimal distance vectors $\mbox{d}\mathbf{r}$ under elastic distortions is given by 
\begin{equation}\label{eq_dr}
\mbox{d}\mathbf{r'}=\frac{\partial\mathbf{r'}}{\partial\mathbf{r}}\cdot \mbox{d}\mathbf{r}
=(\mathbf{I}+\bm{\epsilon}+\mathbf{\Omega})\cdot \mbox{d}\mathbf{r}.
\end{equation}
In this expression, $\mathbf{I}$ is the unity matrix, $\bm{\epsilon}$ denotes the linearized strain tensor obtained as the symmetrized part of the gradient tensor $\nabla\mathbf{u}$ with components $\epsilon_{ij}=[\nabla_ju_i+\nabla_iu_j]/2$, while $\mathbf{\Omega}$ describes rigid rotations and is the antisymmetric part of $\nabla\mathbf{u}$ with components $\Omega_{ij}=[\nabla_ju_i-\nabla_iu_j]/2$. Rigid rotations of the material as a whole do not increase its internal energy. Therefore, only strain deformations $\bm{\epsilon}$ can directly contribute to that part of the generalized energy density $F$ which describes the energetic increase due to elastic deformations. 

\begin{figure}
\centerline{\includegraphics[width=8.5cm]{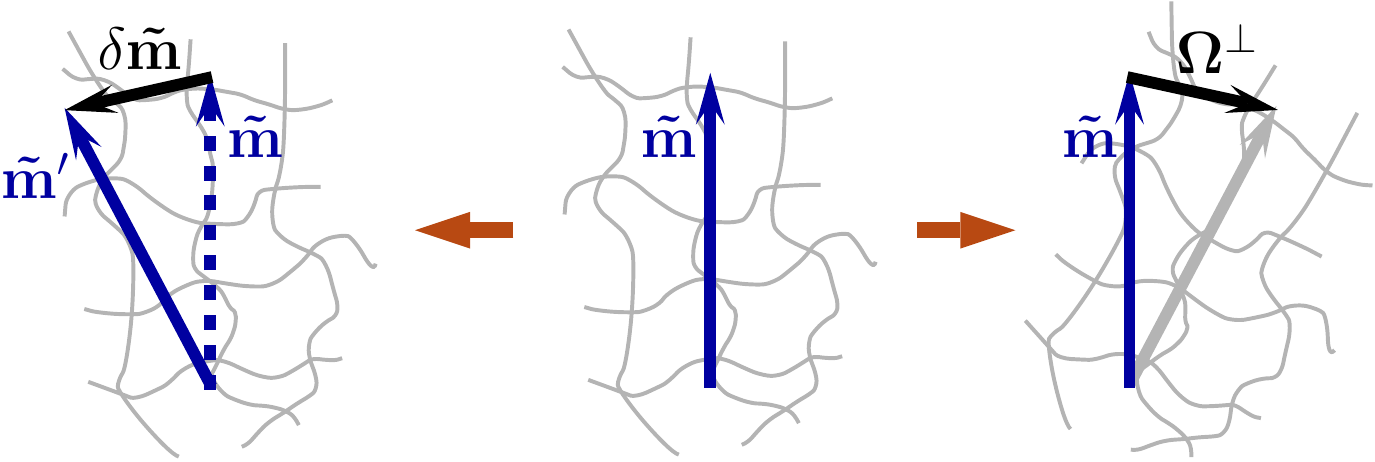}}
\caption{
Schematic illustration of the macroscopic variable of relative rotations. Such relative rotations can occur between the magnetic component, characterized by the anisotropy variable $\mathbf{\tilde{m}}$, and the elastic component, represented by the polymer network indicated in gray. 
On the one hand, if the polymer network remains fixed but the anisotropy axis is reoriented by $\delta\mathbf{\tilde{m}}$ from $\mathbf{\tilde{m}}$ to $\mathbf{\tilde{m}'}$ (center to left), then $\mathbf{\tilde{m}}$ is rotated relatively with respect to the elastic component. On the other hand, the polymer matrix can be rotated relatively with respect to a fixed $\mathbf{\tilde{m}}$ (center to right); the gray arrow indicates how $\mathbf{\tilde{m}}$ would have been reoriented, if it were rigidly anchored within the polymer matrix; this allows to quantify the relative rotation of the polymer network relatively to $\mathbf{\tilde{m}}$ by the vector $\mathbf{\Omega^{\bot}}$. In general, both processes occur simultaneously. The linear variable of relative rotations $\mathbf{\tilde{\Omega}}$ follows as the difference between both reorientations, $\mathbf{\tilde{\Omega}}=\delta\mathbf{\tilde{m}}-\mathbf{\tilde{\Omega}^{\bot}}$, as given by Eq.~(\ref{eq_relrot}). 
}
\label{fig_relrot}
\end{figure}

{Naturally, combining the magnetic component structured in the chain-like aggregates with the elastic component leads to coupling effects.} This implies a further macroscopic variable characterizing the energetic state of the system. As we noted above, due to the particle distribution in the chain-like aggregates, there is an energetic ground state orientation $\mathbf{\tilde{m}}$. {Along $\mathbf{\tilde{m}}$ both the orientation of the structural anisotropy marked by the chain axes and the orientations of the chain magnetizations coincide in the ground state. This is because the magnetic moments of the particles within the same chain on average preferentially orient along the chain axis to minimize the magnetic interaction energy. After completing the crosslinking process during synthesis, the particle positions are fixed with respect to the surrounding polymer matrix. In this indirect way, i.e.~via the memorized particle positions, the chains keep the memory of the ground state orientations of their magnetizations along the chain axes.} Thus it costs energy to rotate the magnetization directions relatively to the structural anisotropy axis memorized by the elastic component. Such relative rotations are included by a vector variable 
\begin{equation}\label{eq_relrot}
\mathbf{\tilde{\Omega}} = \mathbf{\delta\tilde{m}}-\mathbf{\Omega^{\bot}}.
\end{equation}
The first part, $\mathbf{\delta\tilde{m}}=\mathbf{\tilde{m}'}-\mathbf{\tilde{m}}$, directly measures rotations from an initial state $\mathbf{\tilde{m}}$ to a reoriented state $\mathbf{\tilde{m}'}$. If this rotation $\mathbf{\delta\tilde{m}}$ occurs together with a rigid rotation of the whole material, the energy of the system does not change. Thus this reorientation has to be seen relatively to rotations of the embedding polymer matrix, which are described by the second term $\mathbf{\Omega^{\bot}}=\mathbf{\Omega}\cdot\mathbf{\tilde{m}}$ in Eq.~(\ref{eq_relrot}). Only if the two contributions differ from each other has one component been rotated relatively with respect to the other one. 
See Fig.~\ref{fig_relrot} for further illustration. The variables of relative rotations and their consequences were initially introduced and afterwards intensively studied in the context of nematic elastomers, where likewise nematic director rotations can occur relatively to a crosslinked polymer matrix \cite{degennes1980weak,brand1994electrohydrodynamics,muller2005undulation, menzel2006rotatoelectricity,menzel2007cholesteric,menzel2007nonlinear, menzel2008instabilities,menzel2009nonlinear,menzel2009response}. 
Later relative rotations were also included in a macroscopic characterization of active media for the case in which a dynamic preferred direction can rotate relatively to a surrounding gel matrix \cite{brand2011macroscopic}. 

We now have all the variables at hand to reproduce the expression of the generalized energy density $F$ \cite{bohlius2004macroscopic}. This expression corresponds to a systematic expansion around the ground state of the material. It thus only describes small deviations from that ground state. We here confine ourselves to the lowest order of this description, i.e.\ we only consider quadratic terms in the macroscopic variables. As mentioned above, only spatially homogeneous situations are considered for simplicity:   
\begin{eqnarray}
F &=& F_0 + \frac{1}{2}B_iB_i+\frac{\alpha}{2}M_iM_i-M_iB_i \nonumber\\
  & & {}+\frac{1}{2}c_{ijkl}\epsilon_{ij}\epsilon_{kl} \nonumber\\
  & & {}+\frac{1}{2}D_1\tilde{\Omega}_i\tilde{\Omega}_i
        +D_2(\tilde{m}_j\delta_{ik}^{\bot}+\tilde{m}_k\delta_{ij}^{\bot})
            \tilde{\Omega}_i\epsilon_{jk}.
\label{eq_F}
\end{eqnarray}
Here, $\mathbf{B}$ denotes an externally applied magnetic field, and $\mathbf{M}$ is the macroscopic magnetization of the sample. Summation over repeated indices is implied throughout the paper. 

In the first line, as mentioned before, $F_0$ contains all variables that are already present in the characterization of normal fluids and will not be investigated further in the following. {Apart from that, the remaining terms of the first line include energetic contributions due to magnetic interactions. Likewise, they will not be in the focus of the present study as will further be commented on in the next section.} 

The second line of Eq.~(\ref{eq_F}) contains the elastic deformation energy and is quadratic in the strain tensor $\bm{\epsilon}$. Here, the tensor of elastic coefficients $c_{ijkl}$ reflects the macroscopic symmetry of the elastic material. In our case, the systems are uniaxial with the axis of anisotropy given by $\mathbf{\tilde{m}}$. Therefore, we expand \cite{brand1994electrohydrodynamics}
\begin{eqnarray}\label{eq_c}
c_{ijkl} &=& c_1(\delta_{ij}^{\bot}\delta_{kl}^{\bot})
             + c_2(\delta_{ik}^{\bot}\delta_{jl}^{\bot} 
                   +\delta_{il}^{\bot}\delta_{jk}^{\bot}) \nonumber\\
         &&{}+ c_3\tilde{m}_i\tilde{m}_j\tilde{m}_k\tilde{m}_l
             + c_4(\tilde{m}_i\tilde{m}_j\delta_{kl}^{\bot}
                   +\tilde{m}_k\tilde{m}_l\delta_{ij}^{\bot}) \nonumber\\
         &&{}+ c_5(\tilde{m}_i\tilde{m}_k\delta_{jl}^{\bot}
                   +\tilde{m}_i\tilde{m}_l\delta_{jk}^{\bot} \nonumber\\
         &&{}\qquad+\tilde{m}_j\tilde{m}_k\delta_{il}^{\bot}
                   +\tilde{m}_j\tilde{m}_l\delta_{ik}^{\bot}).
\end{eqnarray}
Here, $\mathbf{I}^{\bot}=\mathbf{I}-\mathbf{\tilde{m}}\mathbf{\tilde{m}}$, or in components $\delta^{\bot}_{ij}=\delta_{ij}-\tilde{m}_i\tilde{m}_j$, describes projections into the plane perpendicular to the anisotropy variable $\mathbf{\tilde{m}}$. Expression~(\ref{eq_c}) for the tensor of elastic coefficients is identical to the one given in Ref.~\cite{bohlius2004macroscopic}, with the relations to the coefficients in Ref.~\cite{bohlius2004macroscopic} given by $c_1=\mu_1-\mu_2$ and $c_i=\mu_i$ for $i\in\{2,3,4,5\}$. 

Finally, the last line of Eq.~(\ref{eq_F}) includes the effect of relative rotations. The $D_1$ term states that relative rotations always cost energy. Apart from that, there is a coupling between relative rotations and elastic strain deformations represented by the strain tensor $\bm{\epsilon}$ in the $D_2$ term. This implies that, although relative rotations cost energy, strain deformations can nevertheless excite them under certain conditions. 

We repeat that our description in Eq.~(\ref{eq_F}) represents a lowest order expansion in the macroscopic variables. Only quadratic terms in these variables are included. On the one hand, it is therefore sufficient to insert the linearized versions of the strain tensor $\bm{\epsilon}$ and of the vector of relative rotations $\mathbf{\tilde{\Omega}}$. This maintains the overall quadratic order of the approach. 
On the other hand, we remark that the restriction to this lowest-order quadratic expression is also necessary to keep the present description self-consistent. For example, for the expansion in Eq.~(\ref{eq_c}), we have set the axis of anisotropy $\mathbf{\tilde{m}}$ by the ground state orientation of the chain magnetizations. In reality, however, when the system is not in the ground state any more, the direction of magnetization may deviate from the purely configurational anisotropy axis set by the chain axes. Such deviations are small close to the ground state and lead to higher-order contributions because the generalized energy density in Eq.~(\ref{eq_F}) is already of quadratic order in the strain tensor $\bm{\epsilon}$ and in the vector of relative rotations $\mathbf{\tilde{\Omega}}$. However, if higher-order effects are considered, these deviations would need to be included to obtain a completely consistent description. In principle, one then would need to derive a theory containing both anisotropy orientations, i.e.\ the directions of chain magnetizations and chain axes, separately. 

\section{Mesoscopic model}\label{mesomodel}

In the following we introduce our simplified mesoscopic magnetic particle model, from which we will calculate the macroscopic coefficients in the next section. As mentioned before, the model does not consider microscopic details. For example, single polymer chains of the surrounding elastic matrix are not resolved. The polymer network is rather taken into account in the form of an embedding elastic continuum. 

As a minimum approach, we consider identical magnetic particles. All of them carry the same magnetic dipolar moment of magnitude $m$. Thermal fluctuations are neglected. In a sample of $N$ magnetic particles, this results in a contribution to the system energy per sample volume $V$ of the form 
\begin{equation}\label{eq_Edip}
E_{dip} = \frac{1}{V}\,\frac{\mu_0}{4\pi}\sum_{i<j}^N \frac{(\mathbf{m}_i\cdot\mathbf{m}_j)r_{ij}^2-3(\mathbf{m}_i\cdot\mathbf{r}_{ij})(\mathbf{m}_j\cdot\mathbf{r}_{ij})}{r_{ij}^5}.
\end{equation}
Here the sum runs over all particle pairs. $\mathbf{m}_i$ denotes the magnetic moment of the $i$th particle ($i=1,\dots,N$) with $\|\mathbf{m}_i\|=m$; $\mathbf{r}_{ij}=\mathbf{r}_j-\mathbf{r}_i$ is the distance vector between two particles located at positions $\mathbf{r}_i$ and $\mathbf{r}_j$ ($i,j=1,\dots,N$) with $\|\mathbf{r}_{ij}\|=r_{ij}$. 

\begin{figure}
\centerline{\includegraphics[width=3.cm]{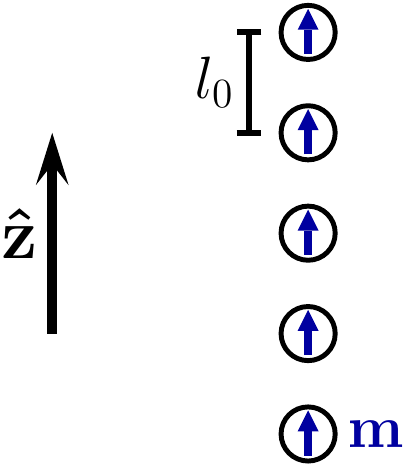}}
\caption{Simplified mesoscopic particle-resolved model. Identical magnetic particles of identical dipolar magnetic moment $\mathbf{m}$ are embedded in a surrounding elastic polymer matrix (not shown). The particles form chain-like aggregates, with the chain axes oriented along the $\mathbf{\hat{z}}$ direction. Within one chain, we assume the same center-to-center distance $l_0$ between all pairs of neighboring particles. 
}
\label{fig_chain}
\end{figure}

We consider regular chain-like aggregates. All chains are assumed to be aligned parallel to the same direction, which can be approximately achieved during synthesis by applying a strong external magnetic field. The center-to-center distance between neighboring particles within one chain is denoted as $l_0$ and assumed to be the same for all such particle pairs, see Fig.~\ref{fig_chain}. 

Both to make progress in the next section and to keep the characterization general and illustrative, we now introduce a row of approximations. 
First, we neglect magnetic interactions between magnetic particles not belonging to the same chain. 
{In other words, we consider a system of sufficiently well-separated magnetic chains within the elastic polymer matrix. This approximation of non-interacting chains is often necessary to make analytical progress \cite{zubarev2013effect}. It appears justified in several cases of real experimental samples featuring magnetic chains of appreciable length \cite{gunther2012xray,borbath2012xmuct}. For such samples, it can be estimated that the magnetic interactions between close particles of neighboring chains should be several hundred times weaker than between nearest-neighbor particles within the same chain. Moreover, in finite-element simulations, inter-chain interactions were found to be negligible concerning the investigated mechanical properties \cite{han2013field}.} 
Second, within the same chain, only nearest-neighbor magnetic interactions are considered. For an infinitely extended chain consisting of an infinite number of particles, this leads to a deviation in the resulting energy by a factor of $\zeta_R(3)\approx1.2$, with $\zeta_R$ the Riemann zeta function. In our minimum model this constitutes a tolerable error. It is then also consistent to neglect interactions between the chains, if the individual chains are separated by a distance large compared to $l_0$. 
As a consequence of these approximations, our description remains more general and does not contain specific configurational properties of the aggregates, which vary from each material to the other due to details of the manufacturing process. 

{We only consider static (or quasistatic) situations. Thus the orientations of the magnetic moments have relaxed to their energetic ground state. In the absence of any misaligning external magnetic field, this is always along the chain axis, even if the chains are distorted by homogeneously being stretched in an oblique direction. For this to be possible, we assume that the magnetic moments can freely reorient to find the energetic minimum, at least in the (quasi)static limit. In other words, we do not consider an explicit permanent orientational memory for the magnetic moments \cite{annunziata2013hardening}. In reality, this can be achieved for example by using small enough particles below a size of about $10$--$15$~nm; in this case the magnetic moments can reorient with respect to the particle axes, which is called N\'{e}el mechanism. Or, for instance, larger spherical particles may be able to rotate as a whole within the pockets of the matrix, at least in the long-time limit. 
}

Apart from that, we assume spatial homogeneity throughout the system. This implies two major approximations. {On the one hand, we assume that the axes of all chains show the same orientation. Within one chain, all dipolar magnetic moments are assumed to point into the same direction.} 
%

{
In reality, two such situations are conceivable. First, one could imagine that the macroscopic magnetization $\mathbf{M}$ of the sample vanishes, $\mathbf{M}=\mathbf{0}$. In the presence of inter-chain interactions, this situation may follow when magnetic moments of different chains point into opposite directions to minimize the overall magnetic energy. On average, there is an equal amount of magnetic moments pointing into the one and into the opposite direction. Then, there is no preferred direction (i.e.\ sense) connected to the macroscopic axis of anisotropy $\mathbf{\tilde{m}}$. Similarly to the case of nematic liquid crystals \cite{degennes1993physics}, the axis of anisotropy cannot distinguish between head or tail, i.e.\ the system from a macroscopic point of view is symmetric under the transformation $\mathbf{\tilde{m}}\leftrightarrow -\mathbf{\tilde{m}}$. Second, the system may feature a net macroscopic magnetization $\mathbf{M}\neq\mathbf{0}$. 
In our picture, this for example can be realized by applying a strong external magnetic field $\mathbf{B}$ of constant magnitude parallel to the chain axes. Then all magnetic moments would point into the same direction, $\mathbf{m}_i=\mathbf{m}$ ($i=1,\dots,N$). 
The total macroscopic magnetization in the sample results as $\mathbf{M}=N\mathbf{m}/V$, with $N$ the total number of particles and $V$ the sample volume. 
The macroscopic variable $\mathbf{\tilde{m}}$ can in this situation be defined as a genuinely polar vector via the macroscopic magnetization, $\mathbf{\tilde{m}}=\mathbf{M}/M$, $M=\|\mathbf{M}\|=(M_iM_i)^{\frac{1}{2}}$. Thus the system in this situation macroscopically shows a head and a tail.}

{
The first case of $\mathbf{M}=\mathbf{0}$ may occur in the absence of an externally applied magnetic field, $\mathbf{B}=\mathbf{0}$. Then the terms after $F_0$ in the first line of Eq.~(\ref{eq_F}) vanish. In the second case of applying a strong external magnetic field $\mathbf{B}\neq\mathbf{0}$ of constant magnitude, all magnetic moments will be homogeneously aligned along the field direction. Then the terms after $F_0$ in the first line of Eq.~(\ref{eq_F}) are constant and can be absorbed into $F_0$. As a consequence, we will not explicitly take these terms into account in the remaining part of this paper.}

{
Remarkably, Eq.~(\ref{eq_F}) was originally derived for a polar variable $\mathbf{\tilde{m}}$ \cite{bohlius2004macroscopic}. However, in the last two lines of this (quasi)static part of the macroscopic characterization, $\mathbf{\tilde{m}}$ only enters to even order [see further Eqs.~(\ref{eq_relrot}) and (\ref{eq_c})]. Therefore, the characterization is equally valid for a non-polar nematic-like variable $\mathbf{\tilde{m}}$. In other words, both of the above cases are covered by the last two lines of Eq.~(\ref{eq_F}). }

On the other hand, our homogeneity assumption implies affine deformations of the polymer matrix. As a consequence, spatially homogeneous strain deformations throughout the sample are assumed, despite the presence of the embedded particles. 
{Naturally, and especially on the length scales of the embedded particles, this is an approximation. Particularly in the close vicinity of the inclusions, the matrix in reality is distorted in an inhomogeneous way. In general, high-resolution finite-element simulation methods are necessary to capture this effect \cite{han2013field,spieler2013xfem}. Strictly speaking, the homogeneity assumption in our analytical treatment is only valid for point-like inclusions.}
The closest situation to match these homogeneity assumptions may be the bulk behavior in a material consisting of chains that on the one hand span the whole sample \cite{gunther2012xray} {and on the other hand are composed of small-sized magnetic particles of sufficiently high magnetic moment}. 

As a result of these approximations, only the volume concentration $c_{nn}$ of nearest-neighboring particle pairs will enter our expressions below. If we denote the number of magnetic particles in a chain by $n$ and the {volume concentration of chains} that consist of $n$ particles by $p(n)$, then the concentration of nearest-neighbors follows as $c_{nn}=\sum_{n=2}^{\infty}(n-1)p(n)$. 

Finally, we assume that the polymer forming the elastic matrix does not interact with magnetic fields. Its magnetic susceptibility is set to zero. Nevertheless, {for $m=0$, thus without magnetic interactions between the embedded particles}, the elastic behavior of the composite material will still be anisotropic. 
This is because of the chain-like particle arrangements in the composite material. 
Within linear elasticity theory, we can thus denote the elastic energy density on the mesoscopic level as
\begin{equation}\label{eq_Eel}
E_{el} = \frac{1}{2}c_{ijkl}^{(0)}\epsilon_{ij}\epsilon_{kl}, 
\end{equation}
where the tensor of elastic coefficients has the same form as in Eq.~(\ref{eq_c}). The elastic coefficients in this state of $m=0$ are denoted as $c_i^{(0)}$ for $i\in\{1,2,3,4,5\}$. These coefficients are used as an input to our mesoscopic model. For illustration and easier traceability of the following arguments, we explicitly expand $c^{(0)}_{ijkl}$ as in Eq.~(\ref{eq_c}) and list the resulting expression of $E_{el}$ for the geometry depicted in Fig.~\ref{fig_chain}, i.e.\ for the chain axes parallel to $\mathbf{\hat{z}}$: 
\begin{eqnarray}
E_{el}(\bm{\epsilon}) & = & \frac{1}{2}c_1^{(0)}(\epsilon_{xx}^2+2\epsilon_{xx}\epsilon_{yy}+\epsilon_{yy}^2) \nonumber\\
&&{}+ c_2^{(0)}(\epsilon_{xx}^2+2\epsilon_{xy}^2+\epsilon_{yy}^2) 
+\frac{1}{2}c_3^{(0)}\epsilon_{zz}^2 
\nonumber\\
&&{}+ c_4^{(0)}(\epsilon_{xx}\epsilon_{zz}+\epsilon_{yy}\epsilon_{zz})
+ 2c_5^{(0)}(\epsilon_{xz}^2+\epsilon_{yz}^2). \nonumber\\ &&
\label{eq_Eel_expl}
\end{eqnarray}

\section{Scale bridging}\label{scalebridging}

In the following, we will derive expressions for the macroscopic coefficients $c_i$ with $i\in\{1,2,3,4,5\}$ in Eq.~(\ref{eq_c}) as well as for $D_1$ and $D_2$ in Eq.~(\ref{eq_F}). These expressions will be determined as a function of the mesoscopic model parameters $m$, $l_0$, $c_{nn}$, and $c_i^{(0)}$ with $i\in\{1,2,3,4,5\}$. We consider the equilibrated state of the uniaxial magnetic gel of $m\neq0$ as the ground state of the system. The generalized energy density in Eq.~(\ref{eq_F}) is considered as an expansion around that ground state. 

\subsection{Stretching deformations and expressions for the elastic coefficients $c_1$, $c_2$, $c_3$, and $c_4$}\label{c1c2c3c4}

{
As mentioned before, in practice the chain-like aggregates are generated by applying an external magnetic field during the final production process of the materials \cite{collin2003frozen,varga2003smart,filipcsei2007magnetic,gunther2012xray, borbath2012xmuct,gundermann2013comparison}. 
In previous studies, it was mentioned that still there can remain finite gaps filled with matrix material between the magnetic particles, and the influence of such gaps was discussed \cite{jolly1996model,coquelle2005magnetostriction}. Especially when using surface functionalized magnetic particles with polymer chains covalently attached to the particle surfaces \cite{frickel2009functional,frickel2011magneto,messing2011cobalt}, a separation by a finite layer of polymer material between neighboring particles is conceivable. For this purpose, in the ideal case, polymer chains should be anchored on the particle surfaces sufficiently early in the process of sample preparation. Particularly, this means before the particles are forced to form the chain-like aggregates. Here, we study the influence on the macroscopic elastic moduli arising from such a finite gap between next-nearest neighboring particles in a chain. 
}

\begin{figure}
\centerline{\includegraphics[width=8.5cm]{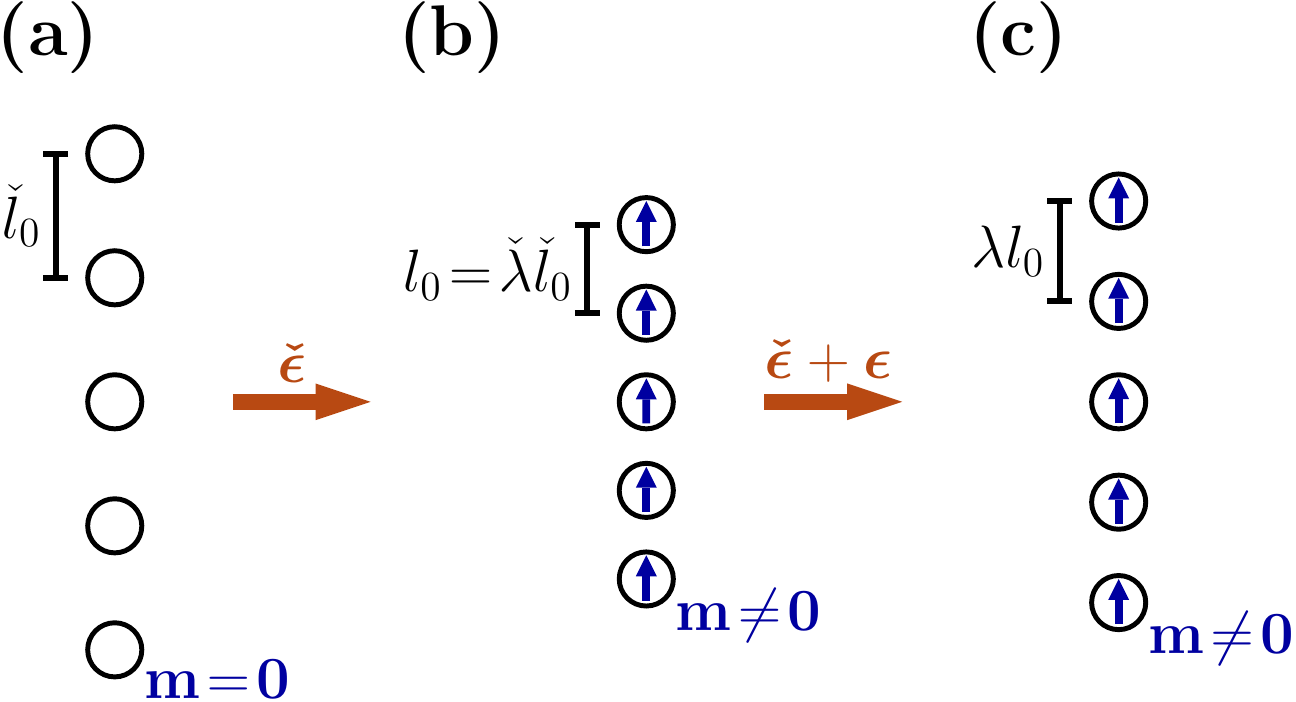}}
\caption{Schematic illustration of the precompressed ground-state in uniaxial magnetic gels. (a) When the particles embedded in the polymer matrix are not magnetized, $\mathbf{m}=\mathbf{0}$, they do not attract each other. The polymer matrix between the particles is in a relaxed state and the particles have a center-to-center distance of $\check{l}_0$. (b) Switching on the magnetic moments of the embedded particles, $\mathbf{m}\neq\mathbf{0}$, the particles attract each other due to the attractive dipolar interactions along the chain axis. The elastic matrix between the particles gets compressed, which leads to a balancing counteracting force. In our assumption of affine deformations, the center-to-center distance between the particles changes from $\check{l}_0$ for $\mathbf{m}=\mathbf{0}$ to now $l_0=\check{\lambda}\check{l}_0$. The corresponding macroscopic deformation is denoted by the strain $\bm{\check{\epsilon}}$. This precompressed state is considered as the ground state of the material in our approach. (c) Additional macroscopic strains $\bm{\epsilon}$ can be imposed onto the material, taking it out of its precompressed ground state and changing the particle center-to-center distance from $l_0$ to $\lambda l_0$. The overall macroscopic strain deformation with respect to the state of $\mathbf{m}=\mathbf{0}$ is $\bm{\check{\epsilon}}+\bm{\epsilon}$; with respect to our precompressed ground state of $\mathbf{m}\neq\mathbf{0}$ it is only $\bm{\epsilon}$. 
}
\label{fig_precompress}
\end{figure}

{
For any non-vanishing magnetic moments $m\neq0$, the particles within a chain attract each other. To maintain a finite separation between the particles, a counteracting repulsive force must balance this attraction. This counteracting force arises from elastic deformations of the matrix material between and around the chain particles. Compared to a (possibly hypothetic) situation of vanishing magnetic moments $m=0$ and relaxed matrix material, the system is compressed along the chain axes when the magnetic moments are non-zero $m\neq0$. See the illustration in Fig.~\ref{fig_precompress}~(a) and (b). Therefore, we call our equilibrated ground state of $m\neq0$ a precompressed state. Then,} 
to derive the correct expressions for the elastic coefficients $c_i$ for $m\neq0$ in our expansion, we must take the precompression into account. 

We start our considerations from the state of $m=0$ in Fig.~\ref{fig_precompress}~(a). In this state, the deformational behavior is described by the elastic coefficients $c_i^{(0)}$. The centers of the magnetic particles are separated by a distance $\check{l}_0>l_0$. With the chain axes along $\mathbf{\hat{z}}$, an axis-symmetric deformation takes place when the magnetic moment switches to non-zero values $m\neq0$: 
\begin{equation}\label{rprime}
\mathbf{r}' = \left(\begin{array}{c}x'\\y'\\z'\end{array}\right) 
= \left(\begin{array}{c}\check{\lambda}_{\bot}x\\\check{\lambda}_{\bot}y\\\check{\lambda} \,z\end{array}\right), 
\end{equation}
with $\check{\lambda}$ and $\check{\lambda}_{\bot}$ the stretch ratios parallel and perpendicular to $\mathbf{\tilde{m}}$, respectively. The strain tensor reads
\begin{equation}\label{eq_checkeps}
\bm{\check{\epsilon}} = \left(\begin{array}{ccc}\check{\lambda}_{\bot}-1 & 0 & 0 \\ 0 & \check{\lambda}_{\bot}-1 & 0 \\ 0 & 0 & \check{\lambda}-1 \end{array}\right). 
\end{equation}

From Eq.~(\ref{eq_Eel_expl}), the resulting elastic deformational energy density can be calculated as
\begin{eqnarray}\label{eq_Eel_epscheck}
{E}_{el}(\bm{\check{\epsilon}}) & = & 
2\left(c_1^{(0)}+c_2^{(0)}\right)(\check{\lambda}_{\bot}-1)^2 + \frac{1}{2}c_3^{(0)}(\check{\lambda}-1)^2 \nonumber\\
&&{}+ 2c_4^{(0)}(\check{\lambda}_{\bot}-1)(\check{\lambda}-1).
\end{eqnarray} 
In addition to that, the energy density of magnetic dipolar interactions results from Eq.~(\ref{eq_Edip}) as
\begin{equation}\label{eq_Edip_lambdacheck}
E_{dip}(\check{\lambda}) \approx -c_{nn}\frac{\mu_0}{2\pi}\frac{m^2}{\check{l}_0^3}\frac{1}{\check{\lambda}^3}.
\end{equation}
Here the approximations discussed in the previous section were applied. The new equilibrium for $m\neq0$ is obtained by balancing the resulting forces, 
\begin{equation}
\frac{\partial(E_{el}+E_{dip})}{\partial\check{\lambda}}=0, \qquad \frac{\partial E_{el}}{\partial\check{\lambda}_{\bot}}=0.
\end{equation}
From Eqs.~(\ref{eq_Eel_epscheck}) and (\ref{eq_Edip_lambdacheck}), this leads to 
\begin{eqnarray}
\check{\epsilon}_{xx}&=&\check{\epsilon}_{yy}= -\frac{1}{2}\,\check{\epsilon}_{zz}\frac{c^{(0)}_4}{c^{(0)}_1+c^{(0)}_2}, 
\label{eq_precompperp}\\
\check{\epsilon}_{zz} &\approx& -c_{nn}\,\frac{3\mu_0}{2\pi}\,\frac{m^2}{l_0^3}\,\frac{c^{(0)}_1+c^{(0)}_2}{\left(c^{(0)}_1+c^{(0)}_2\right)c^{(0)}_3-\left(c^{(0)}_4\right)^2}\quad
\label{eq_precomppar}
\end{eqnarray}
for small enough $|\check{\epsilon}_{zz}|$ to apply the linear elasticity theory and $l_0=\check{\lambda}\check{l}_0$. 

We now are in the precompressed state $\bm{\check{\epsilon}}$ corresponding to Fig.~\ref{fig_precompress}~(b). This is our new ground state. The generalized energy density in Eq.~(\ref{eq_F}) is an expansion around this ground state. To find the corresponding elastic coefficients $c_i$, we consider an additional elastic deformation $\bm{\epsilon}$ as in Eq.~(\ref{eq_F}). Within the framework of linear elasticity theory, the total deformation is obtained as $\bm{\epsilon}^{tot}=\bm{\check{\epsilon}}+\bm{\epsilon}$. 

In the following, we again consider compressional and dilational deformations in the form of a diagonal strain tensor $\bm{\epsilon}$. The system is deformed along $\mathbf{\hat{z}}$ by a stretch ratio $\lambda=1+\epsilon_{zz}$ as depicted in Fig.~\ref{fig_precompress}~(c). This implies changes in both the elastic and the dipolar magnetic energy. We exclude shear deformations, so the strain tensor $\bm{\epsilon}$ is diagonal. The transversal deformations are characterized by the components $\epsilon_{xx}$ and $\epsilon_{yy}$ which only lead to changes in the elastic energy, not in the magnetic interactions. {Moreover, not taking into account shear deformations, we find that the tensor of rigid rotations vanishes, $\mathbf{\Omega}=\mathbf{0}$.} Furthermore, we exclude reorientations of the magnetization, $\delta\mathbf{\tilde{m}}=\mathbf{0}$. Consequently, there are no relative rotations, i.e.\ $\mathbf{\tilde{\Omega}}=\mathbf{0}$. 

On the mesoscopic level, we obtain the changes in the energy density for the magnetic component upon the deformation $\bm{\epsilon}$ from Eq.~(\ref{eq_Edip}) as
\begin{equation}\label{eq_DEdip}
\Delta E_{dip} \approx c_{nn}\,\frac{3\mu_0}{2\pi}\,\frac{m^2}{l_0^3}(\epsilon_{zz}-2\epsilon_{zz}^2)
\end{equation}
and for the elastic component from Eq.~(\ref{eq_Eel_expl}) as
\begin{equation}\label{eq_DeltaEel}
\Delta E_{el} = E_{el}(\bm{\check{\epsilon}}+\bm{\epsilon})-E_{el}(\bm{\check{\epsilon}}).
\end{equation}
The resulting change in the total energy density, $\Delta E=\Delta E_{dip}+\Delta E_{el}$, contains several terms linear in the components of $\bm{\epsilon}$. These terms must vanish when the energy shall represent an expansion around the new precompressed ground state. Indeed, this is the case after inserting Eqs.~(\ref{eq_precompperp}) and (\ref{eq_precomppar}). 

We end up with an expression for the change in the total energy density upon the imposed deformation $\bm{\epsilon}$ that reads
\begin{eqnarray}
\Delta E &=& \frac{1}{2}\Big[c_1^{(0)}(\delta_{ij}^{\bot}\delta_{kl}^{\bot})
             + c_2^{(0)}(\delta_{ik}^{\bot}\delta_{jl}^{\bot} 
                   +\delta_{il}^{\bot}\delta_{jk}^{\bot}) \nonumber\\
         &&{}\quad+ \left(c_3^{(0)}+\Delta c_3\right)\tilde{m}_i\tilde{m}_j\tilde{m}_k\tilde{m}_l \nonumber\\
         &&{}\quad+ c_4^{(0)}(\tilde{m}_i\tilde{m}_j\delta_{kl}^{\bot}
                   +\tilde{m}_k\tilde{m}_l\delta_{ij}^{\bot})\Big]\epsilon_{ij}\epsilon_{kl}.  
\end{eqnarray}
When comparing to the expression for the macroscopic generalized energy density in Eq.~(\ref{eq_F}) together with the expansion in Eq.~(\ref{eq_c}), we find the same structure. Only the third elastic coefficient gets modified. It is the one that directly relates to compressions and dilations along the anisotropy axis $\mathbf{\tilde{m}}$. In our mesoscopic model, this implies compressions and dilations of the magnetic chains, which changes the dipolar magnetic interactions. Compressions and dilations perpendicular to the chain axes leave the structure unchanged. Therefore the corresponding elastic coefficients $c_i^{(0)}$ for $i\in\{1,2,4\}$ remain unaltered. 

As a result, the coefficients in the macroscopic generalized energy density as a function of the mesoscopic model parameters read
\begin{equation}
c_i = c_i^{(0)}, \qquad i\in\{1,2,4\},
\end{equation}
and 
\begin{equation}
c_3 = c_3^{(0)} + \Delta c_3.
\end{equation}
The change in the latter coefficient directly results from changes in the dipolar interaction energy and follows from Eq.~(\ref{eq_DEdip}) as
\begin{equation}\label{eq_Deltac3}
\Delta c_3 \approx -c_{nn}\,\frac{6\mu_0}{\pi}\,\frac{m^2}{l_0^3}.
\end{equation}
We can see that with increasing magnetic moment $m$ the elastic coefficient $c_3$ decreases. This trend has also been observed in finite element simulations in Ref.~\cite{han2013field}. There, it has further been shown that a zig-zag arrangement along the chain, which is not considered here, can lead to the opposite effect. 

We recall that shear deformations have not been considered above. The energetic influence of shear deformations within the plane perpendicular to $\mathbf{\tilde{m}}$, i.e.\ here within the $x$-$y$ plane, is determined by the coefficient $c_2^{(0)}$. See also the coefficient of the term $\epsilon_{xy}^2=\epsilon_{yx}^2$ in Eq.~(\ref{eq_Eel_expl}). Above, we have already found that the corresponding elastic coefficient $c_2^{(0)}$ remains unchanged when $m=0$ is switched to $m\neq0$. This is consistent with the fact that the chain-like structures remain unaltered by shear deformations within the plane perpendicular to $\mathbf{\tilde{m}}$. Furthermore, the precompression does not interfere with these shear deformations because the component $\epsilon_{xy}$ in Eq.~(\ref{eq_Eel_expl}) does not couple to the diagonal components of $\bm{\epsilon}$. 

{None of the above deformations involve a rotation of the magnetic particles different from around the chain axes. Thus our assumption that the magnetic moments can reorient rather freely with respect to the matrix environment does not enter these results. This is different from the next subsection, where shear deformations in a plane containing $\mathbf{\tilde{m}}$ are considered.} Those are given by the components $\epsilon_{xz}=\epsilon_{zx}$ and $\epsilon_{yz}=\epsilon_{zy}$ associated with the elastic coefficient $c_5$. 


\subsection{Shear deformations and expressions for the coefficients $c_5$, $D_1$, and $D_2$}\label{c5D1D2}

\begin{figure}
\centerline{\includegraphics[width=8.5cm]{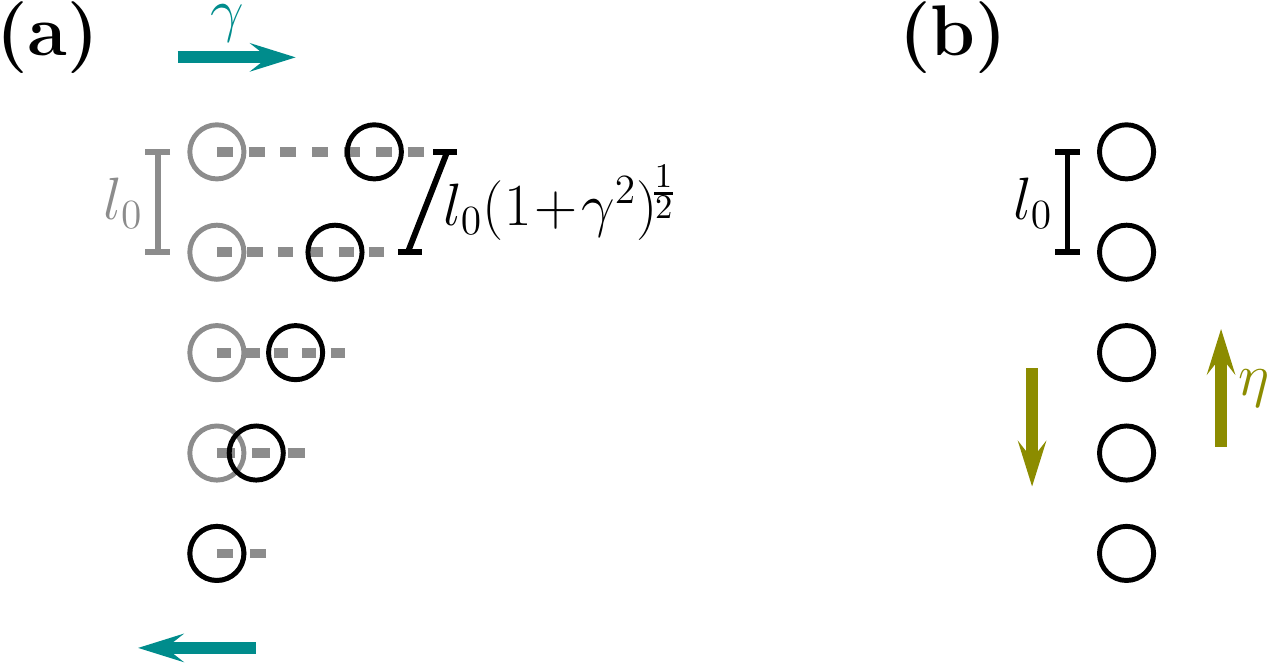}}
\caption{
Illustration of the two kinds of shear deformation of amplitudes $\gamma$ and $\eta$ introduced in Eq.~(\ref{eq_shear}) and their impact on the mesoscopic structures. (a) Applying an affine shear of amplitude $\gamma$ as indicated on the left-hand side results in a reorientation of the chain-like aggregates by affinely horizontally displacing the embedded particles. In addition to that, the center-to-center distances between the particles increase from $l_0$ to $l_0(1+\gamma^2)^{\frac{1}{2}}$. Here, gray spheres show the non-sheared state for $\gamma=0$, whereas black spheres mark the distorted state for $\gamma\neq0$. (b) In contrast to that, the affine shear of amplitude $\eta$ as indicated on the right-hand side does not change the internal structure. 
}
\label{fig_shearpitfall}
\end{figure}

We now consider shear deformations in a plane containing the anisotropy axis $\mathbf{\tilde{m}}$. Without loss of generality, we can assume that the shear is applied in the $x$-$z$ plane. In our mesoscopic set-up, there are two qualitatively different directions in which the shear can be applied. They will be distinguished by the shear amplitudes $\gamma$ and $\eta$ and are indicated in Fig.~\ref{fig_shearpitfall}. 
The corresponding distortion of the material is described by the transformation
\begin{equation}\label{eq_shear}
\mathbf{r}' = \left(\begin{array}{c}x'\\y'\\z'\end{array}\right) 
= \left(\begin{array}{c}{\lambda_{\bot}}x+\gamma z\\ {\lambda_{\bot}}y\\\lambda z+\eta x\end{array}\right). 
\end{equation}
Here, in order not to run into a pitfall resulting from linearized elasticity theory, we introduced again stretch ratios $\lambda$ {and $\lambda_{\bot}$} along {and perpendicular to} the orientation $\mathbf{\tilde{m}}\parallel\mathbf{\hat{z}}$, respectively. The reason is illustrated in Fig.~\ref{fig_shearpitfall}~(a) and explained in the following. 

Simple geometric considerations as indicated in Fig.~\ref{fig_shearpitfall}~(a) show that the shear deformation of amplitude $\gamma$ would result in an increase in the distance between the embedded magnetic particles, from $l_0$ to $l=l_0(1+\gamma^2)^{\frac{1}{2}}$. Consequently, the energy density of magnetic interaction $E_{dip}$ between the dipolar magnetic particles would change. 
{This is a nonlinear effect. The relative length change $(l-l_0)/l_0\approx\frac{1}{2}\gamma^2$ is already of quadratic order in the shear amplitude $\gamma$. Therefore, linear elasticity theory including only linearized expressions for the strain deformations does not cover this effect. We will come back to this issue in the next section, where we demonstrate how this point can be resolved and which consequences this implies. Within our present picture, we circumvent the problem by choosing a deformation that preserves the distances along the chain between the magnetic particles.} 

The value of $\lambda$ from this condition follows as
\begin{equation}\label{eq_lambdacompr}
\lambda = \sqrt{1-\gamma^2} \approx 1-\frac{1}{2}\gamma^2. 
\end{equation}
{When we calculate the corresponding strain tensor following the expressions for the components $\epsilon_{ij}=[\nabla_ju_i+\nabla_iu_j]/2$ within linear elasticity theory and minimize the elastic energy with respect to the deformations in the perpendicular directions $\lambda_{\bot}$, we obtain}
\begin{equation}\label{eq_eps}
\bm{{\epsilon}} = \left(\begin{array}{ccc}
{\left(\frac{c_4}{4(c_1+c_2)}\gamma^2\right)} 
& 0 & \frac{\gamma}{2}+\frac{\eta}{2} \\ 0 & 
{\left(\frac{c_4}{4(c_1+c_2)}\gamma^2\right)} 
& 0 \\ \frac{\gamma}{2}+\frac{\eta}{2} & 0 & \left(-\frac{1}{2}\gamma^2\right) \end{array}\right). 
\end{equation}
{Obviously, the diagonal components describe nonlinear effects $\sim\gamma^2$ and do not enter the expression of our generalized energy density at overall quadratic order. }

Apart from that, together with $\mathbf{\tilde{m}}\parallel\mathbf{\hat{z}}$ we consider 
\begin{equation}\label{eq_delmx}
\delta\mathbf{\tilde{m}}= \left(\begin{array}{c}\delta \tilde{m}_x \\ 0 \\ 0 \end{array}\right), \quad
\mathbf{\tilde{m}}'= \left(\begin{array}{c}\delta \tilde{m}_x \\ 0 \\ 1-\frac{1}{2}(\delta \tilde{m}_x)^2 \end{array}\right), 
\end{equation}
where we, again without loss of generality, only consider deviations $\delta \tilde{m}_x$ in the $\mathbf{\hat{x}}$ direction. Both $\mathbf{\tilde{m}}$ and $\mathbf{\tilde{m}}'$ are unit vectors, which leads to the last entry in $\mathbf{\tilde{m}}'$ truncated after quadratic order in $\delta\tilde{m}_x$. 
Taking into account Eq.~(\ref{eq_relrot}), our set of macroscopic variables is completed by
\begin{equation}\label{eq_Omega}
\mathbf{{\Omega}} = \left(\begin{array}{ccc}0 & 0 & \frac{\gamma}{2}-\frac{\eta}{2} \\ 0 & 0 & 0 \\ \frac{\eta}{2}-\frac{\gamma}{2} & 0 & 0 \end{array}\right), \quad
\mathbf{\tilde{\Omega}} = \left(\begin{array}{c}\delta\tilde{m}_x-\frac{\gamma}{2}+\frac{\eta}{2} \\ 0 \\ 0 \end{array}\right).
\end{equation}
The only elastic coefficient that enters the macroscopic generalized energy density at quadratic order is $c_5$. Due to the underlying mesoscopic effects, it may be altered from $c_5^{(0)}$ by an amount $\Delta c_5$. 

\begin{figure}
\centerline{\includegraphics[width=5.6cm]{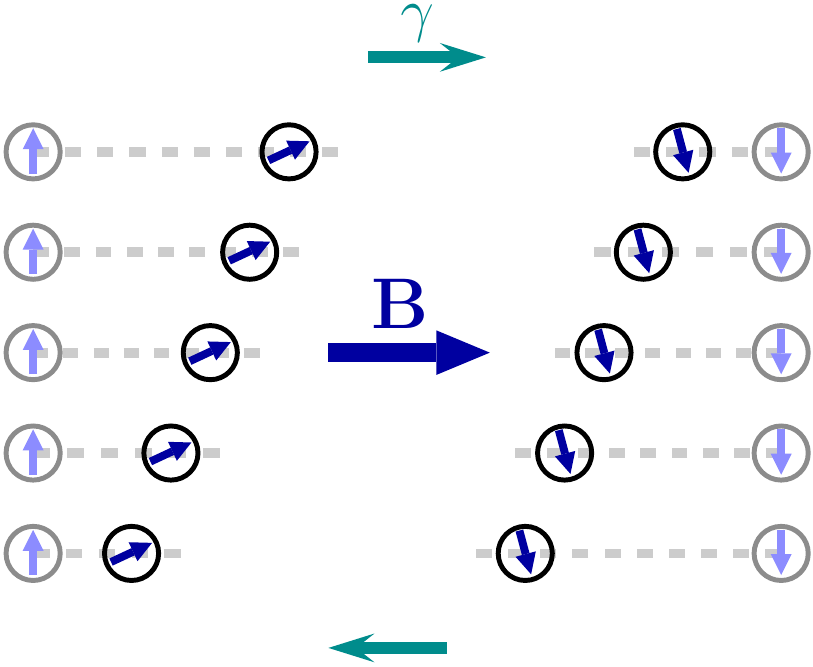}}
\caption{{Inhomogeneous situations induced by an external magnetic field applied perpendicular to the initial chain orientations. Two initially oppositely magnetized parallel chains are indicated by light color on the left- and right-hand sides. Superimposing the magnetic field $\mathbf{B}$ to an affine shear deformation of amplitude $\gamma$ [see also Fig.~\ref{fig_shearpitfall}~(a)] leads to the two different states indicated in the center.}
}
\label{fig_weakB}
\end{figure}

{
On the mesoscopic level and in analogy to Eq.~(\ref{eq_delmx}), we need to include deviations $\delta m_x$ of the magnetic moments from their ground-state directions. For symmetry reasons, it is in the present framework necessary to adhere to the two limiting situations outlined in section \ref{mesomodel}. First, this was the case of vanishing external magnetic field $\mathbf{B}=\mathbf{0}$ and vanishing macroscopic magnetization $\mathbf{M}=\mathbf{0}$, which on the mesoscopic level may be realized by oppositely directed magnetic moments on different chains. Imposing a shear deformation as in Fig.~\ref{fig_shearpitfall}~(a), the magnetic moments will be reoriented to point along the rotated chain axes by energy minimization. This determines $\delta m_x$. Thus the reorientations $\delta m_x$ are dictated by structural reorganization. 
In general, an imposed external magnetic field in this case would imply inhomogeneous situations of the magnetic moment orientations. An illustrative example is depicted in Fig.~\ref{fig_weakB}. The second limiting case was the one of strong external magnetic fields $\mathbf{B}\neq\mathbf{0}$ keeping all magnetic moments directed along itself. In this situation, $\delta m_x$ is imposed by variations of the magnetic field direction. 
}

{It turns out that in the end both limiting cases can be treated by setting $\delta{m}_x=m\,\delta{\tilde{m}}_x$.} On the one hand, for the deformation implied by Eqs.~(\ref{eq_shear}) and (\ref{eq_lambdacompr}), we find the change in the dipolar magnetic energy density from Eq.~(\ref{eq_Edip}) as 
\begin{equation}\label{eq_resultmeso}
\Delta E_{dip} \approx c_{nn}\,\frac{3\mu_0}{4\pi}\,\frac{m^2}{l_0^3}\Big[(\delta \tilde{m}_x)^2 -2\gamma\delta \tilde{m}_x +\gamma^2 \Big]. 
\end{equation}
On the other hand, for the change in the macroscopic generalized energy density, we obtain from Eq.~(\ref{eq_F})
\begin{eqnarray}
\lefteqn{\Delta F - \frac{1}{2}c_5^{(0)}(\gamma^2+2\gamma\eta+\eta^2)} \nonumber\\
&=& \frac{1}{2}D_1\left(\delta\tilde{m}_x\right)^2 + \left(-\frac{1}{2}D_1+D_2\right)\gamma\,\delta\tilde{m}_x \nonumber\\
&&{} +\left(\frac{1}{2}D_1+D_2\right)\eta\,\delta\tilde{m}_x \nonumber\\
&&{} +\left(\frac{1}{8}D_1-\frac{1}{2}D_2+\frac{1}{2}\Delta c_5\right)\gamma^2 \nonumber\\
&&{} +\left(\frac{1}{8}D_1+\frac{1}{2}D_2+\frac{1}{2}\Delta c_5\right)\eta^2 \nonumber\\
&&{} +\left(-\frac{1}{4}D_1+\Delta c_5\right)\gamma\eta.
\label{eq_resultmacro}
\end{eqnarray}
The term headed by the elastic coefficient $c_5^{(0)}$ and shifted to the left-hand side is the bare contribution already present when deforming the anisotropic elastic solid for $m=0$. All contributions on the right-hand side reflect the presence of the magnetic component and its interplay with the elastic matrix. 

We find the values of the macroscopic coefficients $D_1$, $D_2$, and $\Delta c_5$ as a function of the parameters of the mesoscopic structure by comparing the right-hand sides of Eqs.~(\ref{eq_resultmeso}) and (\ref{eq_resultmacro}). Following this procedure term by term, we consistently obtain: 
\begin{eqnarray}
D_1 & \approx & c_{nn}\,\frac{3\mu_0}{2\pi}\,\frac{m^2}{l_0^3}, \label{eq_D1}\\
D_2 & \approx & -c_{nn}\,\frac{3\mu_0}{4\pi}\,\frac{m^2}{l_0^3}, \label{eq_D2}\\
\Delta c_5 & \approx & c_{nn}\,\frac{3\mu_0}{8\pi}\,\frac{m^2}{l_0^3}. \label{eq_Deltac5}
\end{eqnarray}
These equations are the central result of this paper. They represent a bridge from the properties of the mesoscopic structure to expressions for the material parameters in the macroscopic symmetry-based theory. {Within the present linearized analysis, they are independent of the precompression addressed in the previous subsection.}

\subsection{Conditions of thermodynamic stability}\label{thermstab}

To guarantee that the generalized energy density in Eq.~(\ref{eq_F}) provides a well-defined description of real materials, it must be convex in the macroscopic variables. This leads to several conditions for the values of the material parameters. We obtain two sets of variables that decouple from each other, corresponding to our previous sectioning into the subsections \ref{c1c2c3c4} and \ref{c5D1D2}. 

First, as in subsection \ref{c1c2c3c4}, we consider strain deformations that are described by the components of the strain tensor $\epsilon_{xx}$, $\epsilon_{yy}$, $\epsilon_{xy}=\epsilon_{yx}$, and $\epsilon_{zz}$. We have to calculate the Hessian for the corresponding part of the energy density in Eq.~(\ref{eq_F}) and guarantee that it is positive definite. This imposes restrictions on the values of the material parameters. From this procedure, we obtain
\begin{eqnarray}
c_1&>&-c_2, \\ 
c_2&>&0, \\
c_3&>&0 \label{ineq_c3}, \\
c_4^2&<&c_3(c_1+c_2). \label{ineq_c4}
\end{eqnarray}
Since thermodynamic stability must also be guaranteed for the polymer matrix itself for vanishing dipolar interactions, i.e.\ when $m=0$, the same conditions must hold for the coefficients $c_i^{(0)}$ for $i\in\{1,2,3,4\}$. Together with Ineq.~(\ref{ineq_c4}) this guarantees that the precompression in Eq.~(\ref{eq_precomppar}) is well-defined. Combining Eqs.~(\ref{eq_Deltac3}) and (\ref{ineq_c3}), we further find the condition 
\begin{equation}
c_3^{(0)} \gtrapprox c_{nn}\,\frac{6\mu_0}{\pi}\,\frac{m^2}{l_0^3}.
\end{equation}

Second, taking into account the components of the strain tensor $\epsilon_{xz}=\epsilon_{zx}$ and $\epsilon_{yz}=\epsilon_{zy}$, as well as relative rotations $\mathbf{\tilde{\Omega}}$, leads us to conditions for the material parameters discussed in subsection \ref{c5D1D2}. More precisely, we obtain
\begin{eqnarray}
c_5&>&0, \label{ineq_c5}\\
D_1&>&0, \label{ineq_D1}\\
D_2^2&<&c_5D_1. \label{ineq_D2}
\end{eqnarray}
As above, we likewise must demand $c_5^{(0)}>0$. Since $\Delta c_5>0$ in Eq.~(\ref{eq_Deltac5}), there is no further restriction for the coefficient $c_5^{(0)}$. Moreover, our results in Eqs.~(\ref{eq_D1})--(\ref{eq_Deltac5}) do not imply any further restriction for $c_5^{(0)}$ from Ineq.~(\ref{ineq_D2}).

\section{Magnetostrictive effects}\label{magnetostriction}

{
We now turn back to the problem indicated in Fig.~\ref{fig_shearpitfall}~(a) and discussed by Eqs.~(\ref{eq_lambdacompr}) and (\ref{eq_eps}). The shear deformation of amplitude $\gamma$ introduced in Fig.~\ref{fig_shearpitfall}~(a) leads to a change in distance between the magnetic particles within each chain. As explained in Eqs.~(\ref{eq_lambdacompr}) and (\ref{eq_eps}), this is a nonlinear effect of magnitude $\sim\gamma^2$ and thus not covered by the linear elasticity theory. 
}

{
In fact, this change in distance is correctly captured by the expression for the nonlinear strain tensor $\bm{\epsilon}^{nl}$. The components of this strain tensor measure the complete change in distance during the deformation, $(d\mathbf{r}')^2-(d\mathbf{r})^2=2\epsilon^{nl}_{ij}dr_idr_j$. They are given by $\epsilon^{nl}_{ij}=[\nabla_ju_i+\nabla_iu_j+(\nabla_iu_k)(\nabla_ju_k)]/2$ in the Lagrange picture \cite{landau1986elasticity}, to which we adhere for simplicity. 
}

{
We now turn back to the deformation given by Eq.~(\ref{eq_shear}). Including the nonlinear contributions, the strain tensor becomes
\begin{widetext}
\begin{equation}\label{eq_epsnl}
\bm{{\epsilon}}^{nl} = \left(\begin{array}{ccc}
(\lambda_{\bot}-1) + \frac{1}{2}(\lambda_{\bot}-1)^2+\frac{1}{2}\eta^2
& 0 & \frac{\gamma}{2}+\frac{\eta}{2} 
+\frac{(\lambda_{\bot}-1)\gamma+(\lambda-1)\eta}{2}
\\[.1cm] 0 & 
(\lambda_{\bot}-1) + \frac{1}{2}(\lambda_{\bot}-1)^2
& 0 \\[.1cm] 
\frac{\gamma}{2}+\frac{\eta}{2} 
+\frac{(\lambda_{\bot}-1)\gamma+(\lambda-1)\eta}{2}
& 0 & 
(\lambda-1)+\frac{1}{2}(\lambda-1)^2+\frac{1}{2}\gamma^2 \end{array}\right). 
\end{equation}
\end{widetext}
In the last component this tensor correctly captures the nonlinear effect of shear-induced stretching of magnitude $\frac{1}{2}\gamma^2$ along the magnetic chains. 
This is exactly the amount of compression that we had to introduce in the previous section to circumvent this problem and to keep the intra-chain distances constant. 
}

{
It is obvious that the nonlinear parts of the strain tensor get lost in the generalized energy density Eq.~(\ref{eq_F}) to quadratic order. To include them to overall quadratic order, we would need energetic contributions that are linear in the strain tensor. Those are given by the magnetostrictive terms. In the present context, we may denote them as
\begin{equation}\label{magnetostr}
-\frac{1}{2}\zeta_{ijkl}m^2\tilde{m}_i\tilde{m}_j\epsilon_{kl}
\end{equation}
Here, the tensor of magnetostrictive coefficients is called $\bm{\zeta}$ instead of $\bm{\gamma}$ \cite{bohlius2004macroscopic} to avoid confusion with the shear amplitude $\gamma$ introduced in Eq.~(\ref{eq_shear}). 
Due to the symmetry of the system, the tensor $\bm{\zeta}$ can be expanded as
\begin{eqnarray}\label{eq_zeta}
\zeta_{ijkl} &=& \zeta_1(\delta_{ij}^{\bot}\delta_{kl}^{\bot})
             + \zeta_2(\delta_{ik}^{\bot}\delta_{jl}^{\bot} 
                   +\delta_{il}^{\bot}\delta_{jk}^{\bot}) \nonumber\\
         &&{}+ \zeta_3\tilde{m}_i\tilde{m}_j\tilde{m}_k\tilde{m}_l
             + \zeta_4\tilde{m}_i\tilde{m}_j\delta_{kl}^{\bot}
             + \zeta_5\tilde{m}_k\tilde{m}_l\delta_{ij}^{\bot} \nonumber\\
         &&{}+ \zeta_6(\tilde{m}_i\tilde{m}_k\delta_{jl}^{\bot}
                   +\tilde{m}_i\tilde{m}_l\delta_{jk}^{\bot} \nonumber\\
         &&{}\qquad+\tilde{m}_j\tilde{m}_k\delta_{il}^{\bot}
                   +\tilde{m}_j\tilde{m}_l\delta_{ik}^{\bot}).
\end{eqnarray}
}

{
We find that the contributions with the coefficients $\zeta_1$, $\zeta_2$, and $\zeta_5$ are of cubic order. Therefore, they must be neglected in our quadratic expansion. 
Next, the terms with the coefficients $\zeta_3$ and $\zeta_4$ lead to the anticipated contributions linear in $\bm{\epsilon}$. In detail, we obtain $-\frac{1}{2}\zeta_3m^2\epsilon_{zz}$ and $-\frac{1}{2}\zeta_4m^2(\epsilon_{xx}+\epsilon_{yy})$, respectively. 
Finally, the $\zeta_6$ term contains an energetic contribution of quadratic order, namely $-2\zeta_6m^2\mathbf{\tilde{m}}\cdot\bm{\epsilon}\cdot\delta\mathbf{\tilde{m}}$. 
}

{
Now including $\bm{\epsilon}^{nl}$ in the $\zeta_3$ and $\zeta_4$ terms, we performed the same procedure as in section \ref{c5D1D2}. As a result, we obtain the same expressions for $D_1$, $D_2$, and $\Delta c_5$ as before in Eqs.~(\ref{eq_D1})--(\ref{eq_Deltac5}). They are supplemented by
\begin{eqnarray}
\zeta_3 & \approx & -c_{nn}\,\frac{3\mu_0}{\pi}\,\frac{1}{l_0^3}, \label{eq_zeta3}\\
\zeta_4 & \approx & 0, \label{eq_zeta4}\\[.1cm]
\zeta_6 & \approx & 0. \label{eq_zeta6}
\end{eqnarray}
The shift $\Delta c_3$ becomes larger than calculated in Eq.~(\ref{eq_Deltac3}), 
\begin{equation}\label{eq_Deltac3_mod}
\Delta c_3 \approx -c_{nn}\,\frac{15\mu_0}{2\pi}\,\frac{m^2}{l_0^3}.
\end{equation}
}

{
However, there is a major conceptual difference when compared to the previous section. The linear separation of the total strain $\bm{\epsilon}^{tot}$ into a precompression $\bm{\check{\epsilon}}$ and a superimposed strain $\bm{\epsilon}$ via $\bm{\epsilon}^{tot}=\bm{\check{\epsilon}}+\bm{\epsilon}$ is not possible any more when nonlinear contributions to the strain tensor are included. Furthermore, the magnetostrictive terms refer to the state of $m=0$ as an energetic ground state. Therefore, the generalized energy density in Eq.~(\ref{eq_F}) supplemented by the magnetostrictive terms here has to be interpreted as an expansion around the non-precompressed ground state of $m=0$ instead of the precompressed state considered in section \ref{c1c2c3c4}. 
}

{
Finally, we again ask the question of thermodynamic stability. It turns out that via the transformations
\begin{eqnarray}
\bar{\epsilon}_{xx} &=& \epsilon_{xx} + \frac{\zeta_3m^2c_4}{4[c_3(c_1+c_2)-c_4^2]}, \\
\bar{\epsilon}_{yy} &=& \epsilon_{yy} + \frac{\zeta_3m^2c_4}{4[c_3(c_1+c_2)-c_4^2]}, \\
\bar{\epsilon}_{zz} &=& \epsilon_{zz} - \frac{\zeta_3m^2(c_1+c_2)}{2[c_3(c_1+c_2)-c_4^2]} 
\end{eqnarray}
the expression for the generalized energy density again adopts the functional form of Eq.~(\ref{eq_F}). An additional term emerges that, however, is constant for constant $m$. Consequently we obtain the same conditions of thermodynamic stability as in section \ref{thermstab}. The only difference arises from the modified shift $\Delta c_3$ in Eq.~(\ref{eq_Deltac3_mod}), which leads to 
\begin{equation}
c_3^{(0)} \gtrapprox c_{nn}\,\frac{15\mu_0}{2\pi}\,\frac{m^2}{l_0^3}.
\end{equation}
}

\section{Discussion}\label{discussion}

We now briefly discuss some aspects of our results, in particular concerning the model-specific material parameters $D_1$ and $D_2$. 
The first parameter $D_1$ directly measures how difficult it is to perform a relative rotation. Or, in other words, how energetically costly it is to rotate the magnetization directions relatively to the embedding matrix network. This is most easily seen from Eqs.~(\ref{eq_relrot}) and (\ref{eq_F}) when deformations are suppressed, i.e.\ when $\bm{\epsilon}=\mathbf{0}$. Then the structural anisotropy orientation defined by the orientation of the chain-like aggregates remains unaltered. Solely the orientation of the magnetization directions may be changed, for example by applying an additional external magnetic field perpendicular to the chain axes. {Since here $\bm{\epsilon}=\mathbf{0}$ is prescribed, both considered cases, i.e.\ oppositely oriented chain magnetizations on the one hand and aligned magnetic moments throughout the sample on the other hand, can be treated in the same way.} In Eq.~(\ref{eq_resultmacro}) the resulting energetic effect follows only from the first term on the right-hand side. As expected, since $D_1>0$ in Eq.~(\ref{eq_D1}), such relative rotations cost energy. 

\begin{figure}
\centerline{\includegraphics[width=8.5cm]{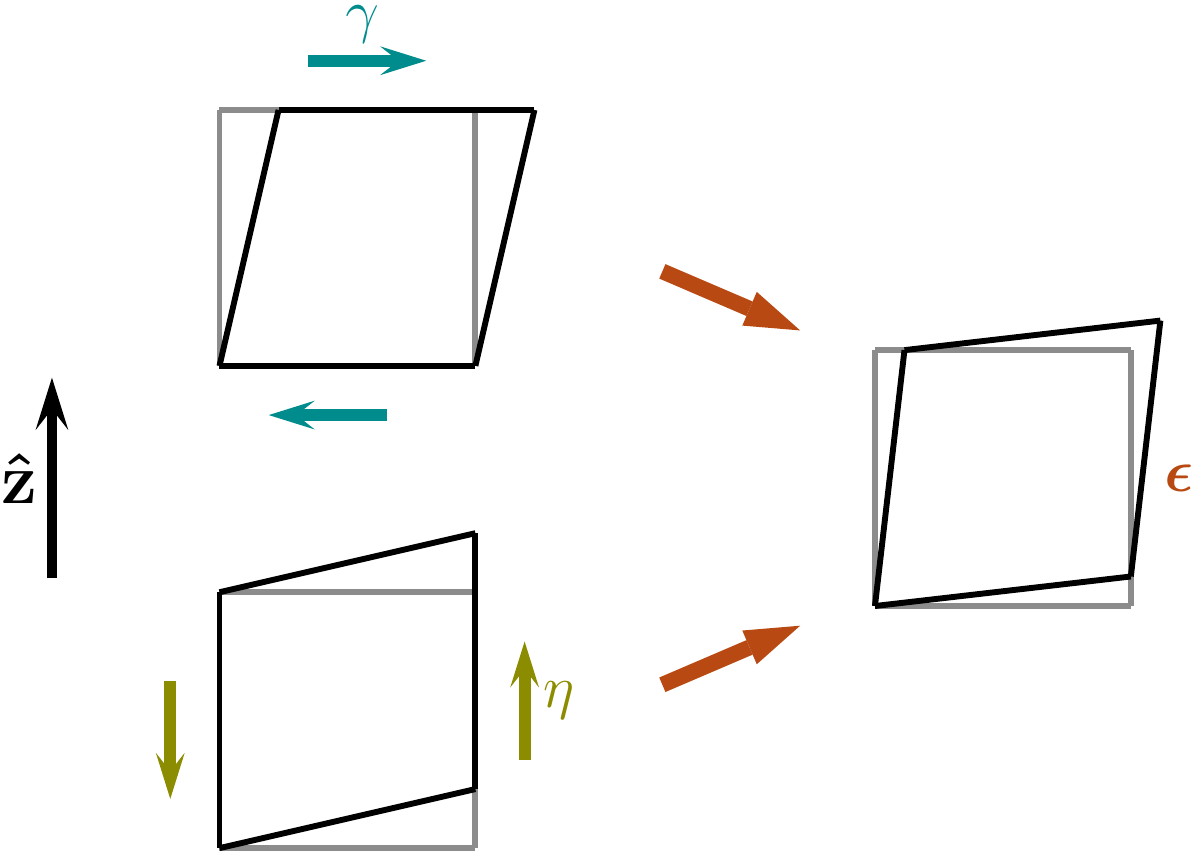}}
\caption{Illustration of the formalism of linear elasticity theory for uniaxial solids in the context of shear deformations. Gray boxes indicate the initial undeformed state, whereas black prisms display the deformed state in each case. In our geometry, the anisotropy axis is initially oriented parallel to the $\mathbf{\hat{z}}$ direction. We introduced in Eq.~(\ref{eq_shear}) two kinds of shear deformation in the plane containing the anisotropy axis: one of amplitude $\gamma$ with displacements perpendicular to $\mathbf{\hat{z}}$; and one of amplitude $\eta$ with displacements parallel to $\mathbf{\hat{z}}$. Linear elasticity theory maps both kinds of shear deformation onto the same components of the strain tensor $\bm{\epsilon}$ [in Eq.~(\ref{eq_eps}) these are the entries $\epsilon_{xz}=\epsilon_{zx}$]. In effect, these components correspond to compressions and dilations along principal axes that are tilted by $45$ degrees with respect to the initial anisotropy axis, as indicated on the right-hand side. 
}
\label{fig_macroepsilon}
\end{figure}

Next, let us focus on the term with the coefficient $D_2$ and try to understand more about the role that it plays in the macroscopic characterization. In some sense, this term is necessary to heal the shortcoming of the remaining part of the macroscopic theory in describing the materials under investigation. We can understand this point by considering the two shear deformations introduced in Eq.~(\ref{eq_shear}) and Fig.~\ref{fig_shearpitfall} with the two amplitudes $\gamma$ and $\eta$. Linear elasticity theory does not distinguish between these two kinds of deformation. Both enter the linearized strain tensor $\bm{\epsilon}$ in the same way at the same positions $\epsilon_{xz}=\epsilon_{zx}$ as can be seen from Eq.~(\ref{eq_eps}). This leads to an identical energetic contribution via the $c_5$ term as given by Eqs.~(\ref{eq_F}) and (\ref{eq_c}). Diagonalizing the linearized strain tensor $\bm{\epsilon}$ from Eq.~(\ref{eq_eps}), we find the orientations of the principal axes of strain for the shear deformations. Within the $x$-$z$ plane, linear elasticity theory maps both kinds of shear deformation onto the same compressive and dilative deformation tilted by $45$ degrees with respect to the anisotropy axis $\mathbf{\tilde{m}}\parallel\mathbf{\hat{z}}$ as depicted in Fig.~\ref{fig_macroepsilon}. 
Moreover, both shear deformations via Eq.~(\ref{eq_Omega}) imply a non-vanishing network rotation, $\mathbf{\Omega}\neq\mathbf{0}$, at least as long as $\gamma\neq\eta$. For simplicity, we consider the case of vanishing reorientations of the magnetization directions, $\delta\mathbf{\tilde{m}}=\mathbf{0}$. {This can be achieved in the situation of a strong external magnetic field aligning the magnetic moments along itself.} From Eq.~(\ref{eq_Omega}), we then always obtain nonzero relative rotations, $\mathbf{\tilde{\Omega}}\neq\mathbf{0}$, as long as $\gamma\neq\eta$. 

\begin{figure}
\centerline{\includegraphics[width=8.cm]{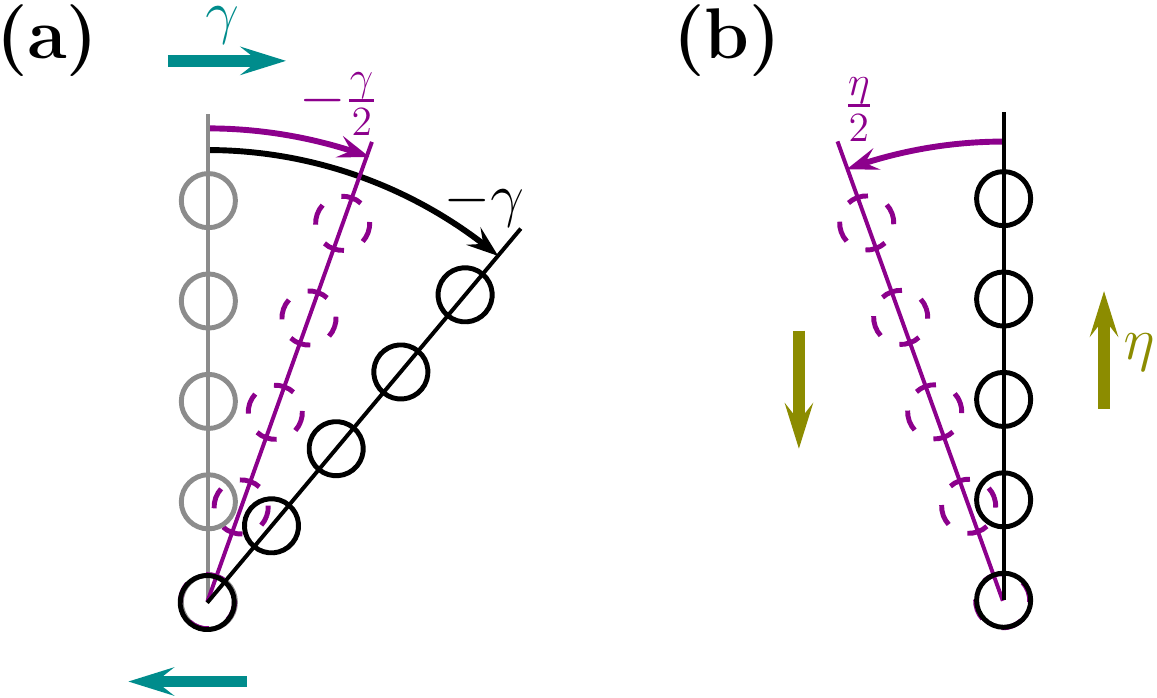}}
\caption{Schematic illustration of the discrepancies arising when the macroscopic level of linear elasticity theory is mapped to the mesoscopic particle-resolved level. Gray color indicates the initial state of the structure in the undeformed state, black color marks the actual reorientation of the chains under affine shear deformations. The broken spheres indicate the reorientations of the structure as they follow from the macroscopic tensor of rigid rotations $\mathbf{\Omega}$ given by Eq.~(\ref{eq_Omega}). For visualization the reorientation angles are strongly exaggerated and generally need to be much smaller in reality to conform to the regime of linear elasticity; distances along the chains are kept constant for illustration, see also the discussion around Eq.~(\ref{eq_lambdacompr}). (a) Under the indicated shear of amplitude $\gamma$, the chain axes during affine transformations are reoriented by an angle $-\gamma$ to linear order. However, the macroscopic theory by the rotation tensor $\mathbf{\Omega}$ only detects reorientations by an angle of half the magnitude, i.e.\ of $-\frac{\gamma}{2}$. (b) Imposing the displayed shear of amplitude $\eta$, the chain axes of the mesoscopic structures are not reoriented at all under affine deformations. The chains as indicated in black remain in their initial orientations. Nevertheless, the macroscopic theory by the rotation tensor $\mathbf{\Omega}$ predicts a rotation of an angle $\frac{\eta}{2}$ in this case. In both cases, the term with the coefficient $D_2$ in the macroscopic generalized energy density can heal these discrepancies and render the approach consistent. 
}
\label{fig_roleD2}
\end{figure}

On the mesoscopic level, a very different picture emerges. Our two kinds of shear deformation of amplitudes $\gamma$ and $\eta$ have qualitatively different effects on the mesoscopic structures, see Fig.~\ref{fig_roleD2}. 
The shear of amplitude $\gamma$ displaces the embedded particles and thus, to linear order, tilts the chain axes by an angle of $-\gamma$ as indicated in Fig.~\ref{fig_roleD2}~(a). However, in the macroscopic description, only a reorientation by an angle $-\frac{\gamma}{2}$ is obtained from the rotation matrix $\mathbf{\Omega}$ in Eq.~(\ref{eq_Omega}). Fixing the magnetization directions in their initial state, $\delta\mathbf{\tilde{m}}=\mathbf{0}$, the macroscopic description thus only predicts half the relative rotation that actually occurs due to structural changes in the sample. In contrast to that, on the mesoscopic level, the orientation of the chain axes remains unaltered under the shear deformation of amplitude $\eta$, see Fig.~\ref{fig_roleD2}~(b). Thus vanishing relative rotations follow on the mesoscopic level when the magnetization directions remain fixed, i.e.\ when $\delta\mathbf{\tilde{m}}=\mathbf{0}$. However, the macroscopic picture again predicts a non-vanishing relative rotation, now of an angle $\frac{\eta}{2}$, resulting from non-vanishing network rotations $\mathbf{\Omega}\neq\mathbf{0}$ in Eq.~(\ref{eq_Omega}). As becomes obvious from this discussion, the discrepancies follow from the use of the symmetrized strain tensor $\bm{\epsilon}$ and the antisymmetric rotation matrix $\mathbf{\Omega}$ in the macroscopic theory, which mixes the effects of the two different kinds of shear deformation. The $D_2$ term removes the resulting discrepancies and restores the compatibility between the macroscopic and mesoscopic approaches. We have demonstrated this fact by showing that the macroscopic expression for changes in the generalized energy density in Eq.~(\ref{eq_resultmacro}) can be consistently mapped onto the energetic changes obtained on the mesoscopic level in Eq.~(\ref{eq_resultmeso}) by choosing an appropriate value of the coefficient $D_2$ in Eq.~(\ref{eq_D2}). Therefore, the term with the coefficient $D_2$ does not only include additional effects into the macroscopic theory. It rather can be viewed as a central ingredient to render the theory consistent. 

We end this discussion by some more remarks about the $D_2$ term, stressing the very interesting observation in Eq.~(\ref{eq_D2}) that $D_2<0$. It is at this point instructive to compare to the situation of nematic liquid crystalline elastomers \cite{brand2011physical}. The macroscopic symmetry-based generalized energy density describing these materials contains a part that is formally identical to the one described here. It is the same part that we put into the focus of this manuscript, i.e.\ the last two lines of the expression in Eq.~(\ref{eq_F}). For nematic liquid crystalline elastomers, it was found that a negative coefficient $D_2$ leads to a reorientation of the anisotropy axis when the material is stretched perpendicularly to its initial orientation. In other words, the anisotropy axis reorients towards the stretching direction. This is connected to a pronounced nonlinearity in the corresponding stress-strain curves. To detect these properties, one has to extend the investigation to the nonlinear regime \cite{urayama2007stretching,menzel2009nonlinear,menzel2009response}. Based on our scale-bridging result, a similar behavior may also be observed for the described uniaxial magnetic gels. This issue should be further tested also by future experimental investigations.

\section{Conclusions}\label{conclusions}

In this work, we established a bridge from a simple mesoscopic model to the macroscopic symmetry-based continuum description of uniaxial magnetic gels. On the mesoscopic level, we considered chain-like structures of dipolar magnetic particles embedded in an elastic polymer matrix. Several approximations were introduced to allow for the scale-bridging procedure and keep the description general. Among them, we assumed affine deformations of the polymer matrix and spatial homogeneity, we neglected all magnetic interactions except for nearest-neighbor ones, and we neglected all inter-chain interactions. In this way, we could explicitly determine expressions for the macroscopic material parameters as a function of the mesoscopic model parameters. On the one hand, we could demonstrate that the macroscopic theory provides a concise description of the materials within the framework of our simplified mesoscopic model. On the other hand, the mesoscopic model can serve to provide an illustrative background for the processes occurring in the materials. In particular, the meaning of the different terms in the macroscopic theory becomes more concrete and can be connected to a descriptive picture. 

Understanding how the structural processes and properties on the particle level can influence the overall material behavior is an important ingredient to design composite materials of optimal and desired features. Moreover, knowing the values of the macroscopic material parameters, qualitative predictions for the macroscopic behavior and properties can be made. We hope that such predictions will also stimulate further experimental investigations.  

Naturally, in the future the presented approach can be refined by more realistic input on the mesoscopic level. Such attempts should best be geared to real samples and materials analyzed by experimental techniques. For instance, three-dimensional particle distributions could be determined by x-ray microtomography and extracted by image analysis \cite{gunther2012xray,borbath2012xmuct,gundermann2013comparison,tarama2014tunable,pessot2014structural}. Of course, the results then become specific to one particular material or sample. 

Finally, our bridge between different length scales started on the mesoscopic level that resolved the individual magnetic particles. We still treated the elastic polymer matrix as an elastic continuum. Our bridge could be extended further down to more microscopic levels, where also individual polymer chains are resolved. In this way, the coefficients on the mesoscopic level could be determined from the microscopic model parameters. A corresponding initial attempt along these lines is currently in progress \cite{pessot2014holm}, drawing a first simplified realization of scale-bridging from the microscopic up to the macroscopic description within reach.

\begin{acknowledgments}
The author thanks the Deutsche Forschungsgemeinschaft for support of this work through the priority program SPP 1681. 
\end{acknowledgments}


\end{document}